\documentclass{article}
\usepackage[T1]{fontenc} 
\usepackage[utf8]{inputenc} 
\usepackage{ismir,amsmath,cite,url}
\usepackage{graphicx}
\usepackage{color}

\usepackage{tabularx, makecell,array,booktabs, multirow}
\newcolumntype{P}[1]{>{\centering\arraybackslash}p{#1}}

\newcommand{\fref}[1]{Figure~\ref{#1}}
\newcommand{\tref}[1]{Table~\ref{#1}}
\newcommand{\sref}[1]{Section~\ref{#1}}

\usepackage{lineno}

\title{Few-shot Drum Transcription in Polyphonic Music}

\multauthor
{Yu Wang$^1$ \hspace{0.2cm} Justin Salamon$^2$ \hspace{0.2cm} Mark Cartwright$^1$ \hspace{0.2cm} Nicholas J. Bryan$^2$ \hspace{0.2cm} Juan Pablo Bello$^1$} {$^1$ Music and Audio Research Lab, New York University, USA\\
$^2$ Adobe Research, San Francisco, USA\\
{\tt\small \{wangyu, mark.cartwright, jpbello\}@nyu.edu, \{salamon, nibryan\}@adobe.com}
}
\def\authorname{Y. Wang, J. Salamon, M. Cartwright, N. J. Bryan, and J. P. Bello}

\begin{document}

\maketitle
\begin{abstract}
Data-driven approaches to automatic drum transcription (ADT) are often limited to a predefined, small vocabulary of percussion instrument classes. Such models cannot recognize out-of-vocabulary classes nor are they able to adapt to finer-grained vocabularies. In this work, we address open vocabulary ADT by introducing few-shot learning to the task. We train a Prototypical Network on a synthetic dataset and evaluate the model on multiple real-world ADT datasets with polyphonic accompaniment. We show that, given just a handful of selected examples at inference time, we can match and in some cases outperform a state-of-the-art supervised ADT approach under a fixed vocabulary setting. At the same time, we show that our model can successfully generalize to finer-grained or extended vocabularies unseen during training, a scenario where supervised approaches cannot operate at all. We provide a detailed analysis of our experimental results, including a breakdown of performance by sound class and by polyphony.
\end{abstract}

\section{Introduction}\label{sec:introduction}
Automatic Drum Transcription (ADT) aims at deriving a symbolic annotation of percussion instrument events from a music audio recording. It is a subtask of Automatic Music Transcription, where the aim is to transcribe all events within a musical piece. An accurate ADT system enables diverse applications in music education, music production, music search and recommendation, and computational musicology.

Early studies on ADT often combined multiple signal processing, information retrieval, and machine learning techniques such as support vector machines (SVM) and hidden Markov models (HMM) \cite{gillet2004, tzanetakis2005, fitzgerald2002}. While these methods work well when applied to solo drum recordings, they often generalize poorly when applied to polyphonic music \cite{gillet2005}. Recent approaches utilizing non-negative matrix factorization (NMF) \cite{Wu2015, wu2017} and deep neural networks \cite{vogl2017, southall2017cnn} have shown promising performance in the presence of polyphonic music. However, such systems are often limited to transcribing a very small subset of percussive sound classes, such as the bass drum (BD), snare drum (SD), and hi-hat (HH). For deep learning methods, in particular, this is mainly due to the limited number and size of ADT datasets, and the small class vocabulary size of the annotations in these datasets \cite{gillet2006, southall2017, vogl2018, dittmar14}. Recently, studies have utilized synthetic data and deep learning to expand ADT systems to support transcribing larger vocabularies of 10 or more instruments \cite{vogl2018, cartwright2018}. However, 10--20 classes are still far from the wide gamut of percussive instruments used in recorded music. For example, rare or non-western percussion sounds are usually considered out-of-vocabulary. Moreover, when transcribing different datasets, we often need to manually map the percussion instruments in a dataset to the limited output vocabulary of an existing ADT system with reduced granularity. It can also be challenging for ADT systems that utilize fully-supervised learning to generalize to different musical genres or diverse drum sounds \cite{vogl2018}.

\begin{figure}[t]
\centering\includegraphics[trim={8cm 20.5cm 12cm 3cm},clip, width=\columnwidth]{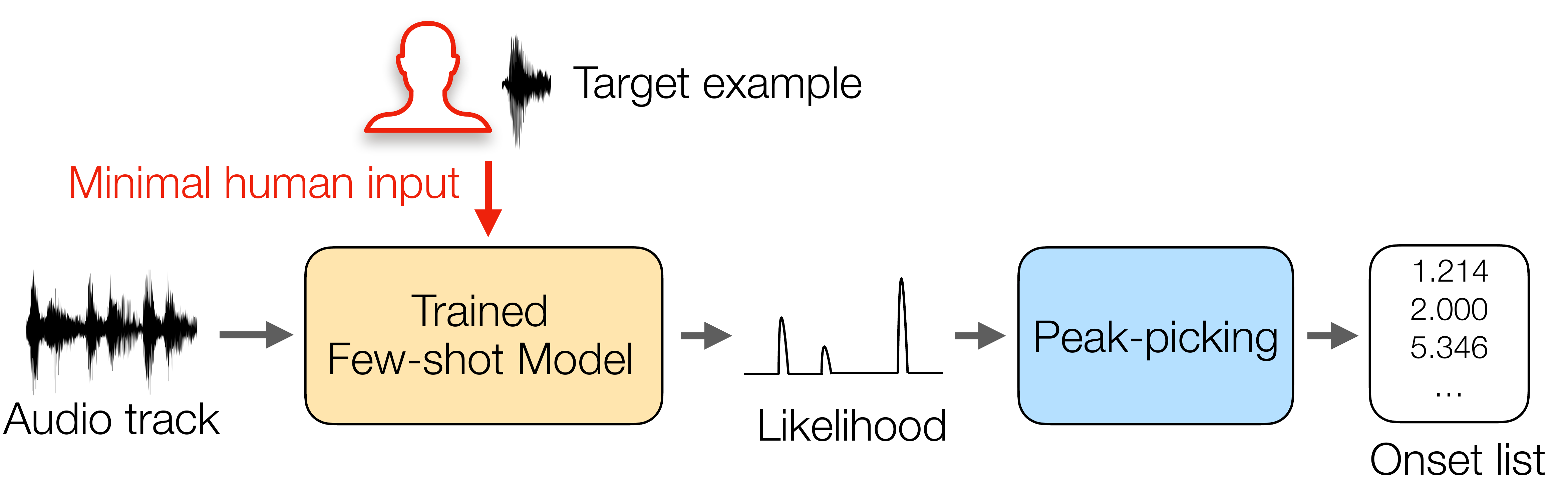}
\caption{Few-shot drum transcription. We input a music recording and one or more target example sounds into our trained model, output the likelihood of the selected event over time, and then post-process to generate onset times per percussion instrument.}
\label{fig:few-shot-adt}
\end{figure}

Recently, few-shot learning has been proposed for recognizing and detecting novel sound events \cite{wang2020, chou2019, cheng2019, shi2020}, which is of great relevance to ADT. Under this paradigm, a model is trained to learn to recognize novel classes, unseen during training, given only very few examples from each class at inference time. Expanding on this, a few-shot sound event detection system for open vocabularies was recently proposed \cite{wang2020}, yielding a search-by-example paradigm where a human first selects a handful of target example sounds that are passed to a trained model that automatically locates similar sounding events within the same audio track. This work, however, was developed for speech data, while other studies have focused on environmental sound. The main challenges in applying few-shot learning to music audio are the limited size of available datasets and the polyphonic nature of music audio. While few-shot learning methods are designed to work with few labeled examples at inference time, they still require large amounts of labeled data at train time. Standard few-shot models are designed to solve multi-class problems where only one class is active at a time, while polyphonic music is inherently multi-label since multiple instruments can be active at once.  


In this work, we propose a new paradigm for drum transcription in polyphonic music by introducing few-shot learning. Instead of trying to train a standard supervised model with more data for better generalizability, we propose to incorporate minimal human input with a few-shot model. Our proposed few-shot drum transcription system is shown in \fref{fig:few-shot-adt}, which can easily adapt to detecting a novel percussion instrument given a handful (e.g., five) of labeled target examples. By doing so, we can support open-vocabulary drum transcription, while minimizing the human labeling effort in order to make the transcription process as close to automatic as possible. To address the aforementioned challenges to applying few-shot learning to polyphonic music audio, we propose (1) utilizing a large synthetic dataset for training and (2) transcribing one percussion instrument at a time as a binary classification problem during inference. We evaluate our proposed model on multiple real music ADT datasets, compare it to a state-of-the-art supervised learning benchmark, and provide a detailed analysis of our model's performance including breakdowns by instrument class and polyphony. We show that our approach not only matches or outperforms past methods, but enables open vocabulary drum transcription, which is highly-advantageous for real-world applications.
\section{Methods}
\subsection{Prototypical Networks with Episodic Training}
In our work, we focus on metric learning-based few-shot methods and, in particular, prototypical networks \cite{Koch2015, vinyals2016, snell2017, sung2018}. Prototypical networks have been found to perform well on several audio-related tasks \cite{pons2019, wang2020, chou2019, shi2020}, rely on a simple training framework, and support efficient feed-forward inference \cite{snell2017}, all of which are advantageous for our problem domain. They are designed to project an input audio example into a discriminative embedding space such that similar sounding events are clustered around a single prototype (average class embedding) via a neural network. Classification is then performed for an embedded query point by simply finding the nearest class prototype via the squared Euclidean distance.

Few-shot learning models, and prototypical networks specifically, are typically trained to solve a \textit{C-way K-shot} multi-class classification task. In this setup, the method is tasked with labeling a query recording with one of $C$ novel class labels, given $K$ labeled examples per class at inference time, where $K$ is typically a small number in the range of one to five. The availability of only very few examples of the new classes limits our ability to fine-tune a pre-trained model. To address this, \textit{episodic training} has been proposed to train a prototypical network, which mimics the few-shot inference problem during training, improving model generalizability \cite{vinyals2016}. In each training iteration, a training \textit{episode} is formed by randomly selecting $C$ classes from the training set. For each selected class, $K$ samples are first selected to build a support set $\mathcal{S}$ of size of $C \times K$, while a disjoint set of $q$ samples are selected to form a query set $\mathcal{Q}$ of size $C \times q$. Prototypes $\mathcal{M} = \{\mu_{1}, ..., \mu_{C}\}$ are the mean vectors of the embedded support samples belonging to each class:
\[
\mu_{c} = \frac{1}{K} \sum_{(x,y)\in \mathcal{S}_{c}} f_\theta (x), 
\]
where $\mathcal{S}_{c}$ denotes a set of examples labeled with class $c$ and $f_\theta$ is parametrized by a neural network. Given a sample $\mathbf{x}_{q}$ in $\mathcal{Q}$, we take a softmax over distances to the prototypes in the embedding space to obtain per-class likelihoods:  
\[
p_{\theta}(y=c \mid \mathbf{x}_{q}) = \frac{\exp(-d(f_{\theta}(\mathbf{x}_{q}), \mu_{c}))}{\sum_{c'}\exp(-d(f_{\theta}(\mathbf{x}_{q}, \mu_{c'}))},
\]
where $d$ is the squared Euclidean distance function. The training objective is to minimize the negative log-likelihood of the true class $c$:
\[
\mathcal{L(\theta)} = -\log p_{\theta}(y=c\mid \mathbf{x}_{q}).
\]
Therefore, in each training episode, the model is learning to solve a \textit{C-way K-shot} classification task. By training with a large collection of episodes, each consisting of a different set of $C$ classes, the model learns \emph{how to learn} from limited labeled data and obtains a class-agnostic discriminative ability. In this work, we train a prototypical network on a \textit{10-way 5-shot} classification task \cite{wang2020} as the few-shot model in our proposed system. 

\subsection{Few-shot Drum Transcription}
\label{few-shot-dt}
While the training task is a specific \textit{C-way K-shot} classification, the trained few-shot model provides an embedding function that projects the input data into a discriminative space in which sound events are classified by finding the nearest class prototype, where each prototype is derived from a few examples. We propose to use this embedding space for percussion sound event detection by providing a support set containing examples for both the positive (target) and negative (non-target) classes, and classify a given query by measuring its distance to the positive and negative class prototypes. Here, the trained few-shot model is essentially performing a binary, 2-way, classification at inference time. 

Given a target instrument and an audio track, we first slice the track into a series of query frames. To construct a support set of labeled examples for the few-shot model, we randomly sample target examples from the track as positive examples, simulating the human input in \fref{fig:few-shot-adt}, and take all frames within the track as negative examples to model the non-target class. Note that, while the full track will also contain the target class, previous work has shown that since the target class is relatively sparse compared to the full track, this strategy works well \cite{wang2020}. Given the support set, the trained few-shot model outputs the likelihood of each query frame containing the target class. Finally, we perform peak-picking on these probabilities to get a list of onset locations as is commonly done for ADT \cite{vogl2018, cartwright2018}. 

\section{Experimental Design}
To evaluate the proposed few-shot drum transcription paradigm, we first train a prototypical network as our few-shot model on a large synthetic dataset. Then, we apply the trained model to three real-music ADT datasets to get transcription performance. We focus on one target instrument at a time and use randomly selected target examples to simulate human input at inference time.   

\subsection{Dataset for Episodic Training: Slakh2100}
We use the Slakh2100 dataset to train our few-shot model \cite{manilow2019}. Slakh2100 is synthesized from the Lakh MIDI Dataset \cite{raffel2016} using professional-grade sample-based virtual instruments. It contains 2100 automatically mixed tracks and accompanying MIDI files, totaling 145 hours of mixtures. Slakh2100 is synthesized using eight different drum patches, where each patch can be viewed as a unique drummer playing a unique drum kit. In each patch, there are around 25 to 45 different percussion classes, each consisting of a combination of a percussion instrument with a playing technique. For episodic training, we alternatively define a class as a specific percussion class (e.g. snare drum side stick) played by a specific patch, resulting in a total of 282 classes. Each drum patch has its own MIDI note-instrument mapping, which does not follow the general MIDI convention. We manually check the mapping to group duplicates and remove empty ones. Each patch is used to synthesize approximately 250 songs. We partition the dataset into patch-conditional train, validation and test splits using 5, 1, 2 patches per split, respectively.  

\subsection{Evaluation Datasets}
\subsubsection{ENST-Drums}
ENST-Drums is a dataset of recordings from three drummers each playing a different drum kit \cite{gillet2006}. It contains drum onset annotations for 20 classes of percussion sounds. While the dataset also contains many solo drum recordings, we only use the subset of 64 recordings with accompaniment for evaluation. The accompaniments are mixed with corresponding drum tracks using a scaling factor of 1/3 and 2/3 to get natural-sounding mixtures and to be consistent with prior studies \cite{gillet2005waspaa, wu2017,vogl2018}.

\subsubsection{MDB-Drums}
MDB-Drums is a set of 23 fully-produced music tracks from the MedleyDB dataset \cite{southall2017, bittner2014}. It contains two levels of drum onset annotations --- we use the finer level which divides the classes by instrument and playing technique, resulting in 21 classes. 

\subsubsection{RBMA13}
RBMA13 consists of 30 fully-produced music tracks in the genres of electronic dance music, singer-songwriter, and fusion-jazz \cite{vogl2018}. The drum sounds of this set are more diverse compared to the previous sets, and it is considered a particularly difficult dataset \cite{vogl2018}. It contains annotations for 23 percussive classes. 

\subsection{Training}
For each percussion instrument onset in Slakh2100, we center a 250 ms context window around the onset as the input to the model. 
We compute a log-scaled Mel-spectrogram from the context window with \texttt{librosa} \cite{mcfee2019} using a window size of 46 ms (2048 samples for a sample rate of 44.1 kHz) and a hop size of 10 ms. In preliminary experiments, we studied a range of short (160 ms) to long (500 ms) context windows and found that a 250 ms window yields consistent, well-performing results across different datasets. We conjecture that a 250 ms window is wide enough to capture most percussive onsets, while also capturing some context around the onset. 

To construct a \emph{10-way 5-shot} training episode, we randomly sample a drum patch from the training set, sample 10 percussion instrument classes from the drum patch, and sample 5 instances per class as the support set. Note that, while each instance is guaranteed to contain the target class, it may also contain other sound classes if they overlap in time with the target class. The query set is comprised of 16 separate instances per each of the 10 classes \cite{chen2019}.

\begin{figure}[t]
\centering\includegraphics[trim={14cm 14cm 19cm 2cm}, clip, width=\columnwidth]{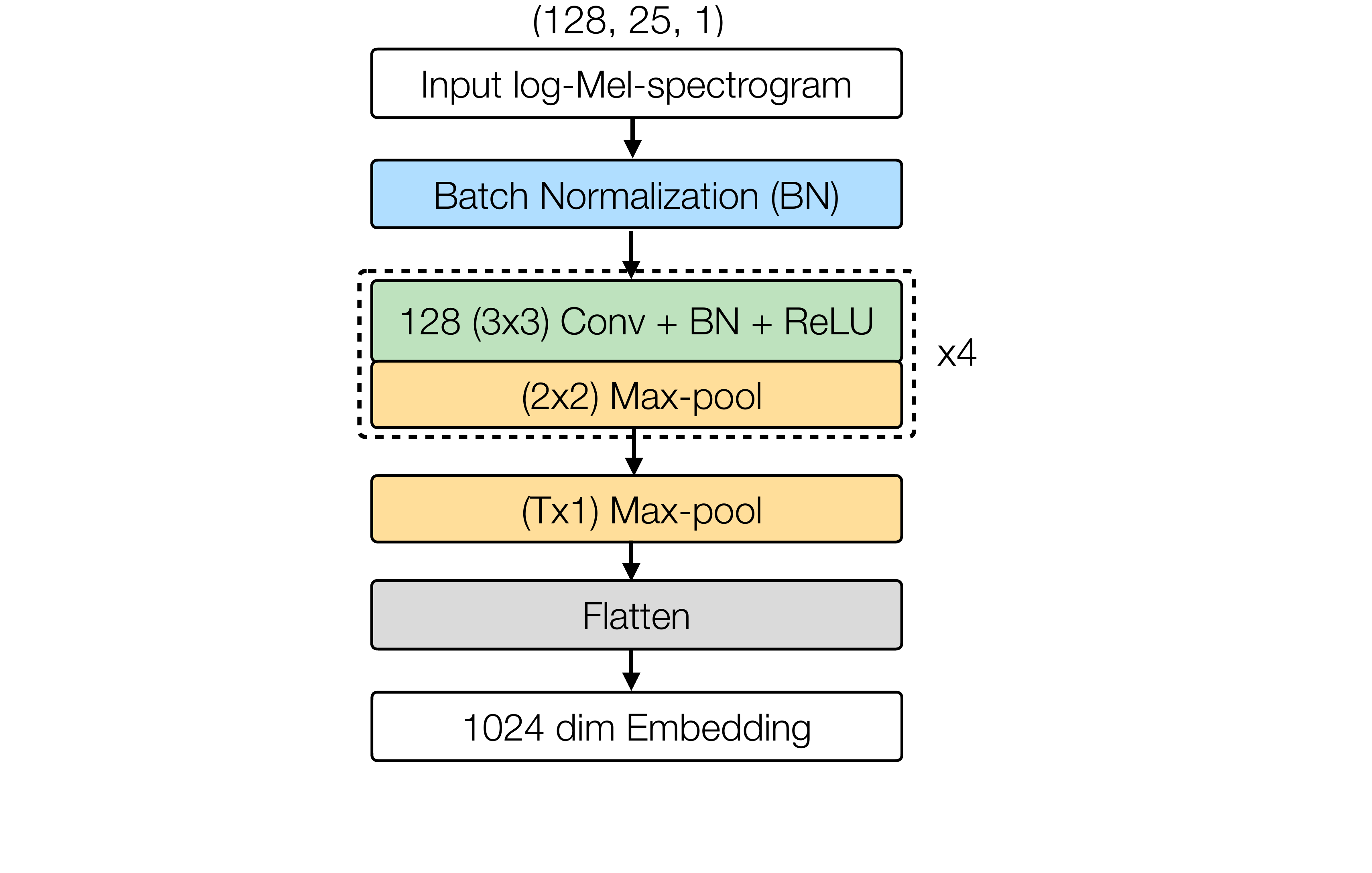}
\vspace{-10mm}
\caption{Our backbone prototypical network embedding model architecture.}
\label{fig:emb}
\end{figure}

\begin{table*}[t]
\centering
\small
\begin{tabular}{P{4.3em}P{4.7em}P{6.7em}|P{2.3em}P{2em}P{2.5em} P{2.3em}P{2em}P{2.5em} P{2.3em}P{2em}P{2.5em}}
\toprule
Vocab size & Model & Target examples & \multicolumn{3}{c}{ENST-Drums} & \multicolumn{3}{c}{MDB-Drums} & \multicolumn{3}{c}{RBMA13}  \\
\cmidrule(lr){4-6} \cmidrule(lr){7-9} \cmidrule(lr){10-12}
 & & & micro & macro & macro* & micro & macro & macro* & micro & macro & macro*
\\
\midrule
 18 &  CRNN \cite{vogl2018} & - & \textbf{0.67}  & - & 0.74   & \textbf{0.60} & - & 0.78  & 0.47  & - &  0.64 \\
\cmidrule(lr){2-12}
 & Proto. Net & Include & 0.55  & \textbf{0.54} & \textbf{0.80}   & 0.58 & \textbf{0.60} &  \textbf{0.87} & \textbf{0.56}  & \textbf{0.55} & \textbf{0.81}  \\
   &  & Exclude & 0.49  & 0.45  & 0.76  & 0.51 & 0.49 & 0.83  & 0.54  & 0.50 & 0.79  \\
\bottomrule
\end{tabular}
\caption{F-measure evaluated on real polyphonic music datasets under a fixed 18-class vocabulary \cite{vogl2018}.}
\label{tab:fix-vocab}
\end{table*}

\begin{table*}[t]
\centering
\small
\begin{tabular}{P{4.3em}P{4.7em}P{6.7em}|P{2.3em}P{2em}P{2.5em} P{2.3em}P{2em}P{2.5em} P{2.3em}P{2em}P{2.5em}}
\toprule
Vocab size & Model & Target examples & \multicolumn{3}{c}{ENST-Drums} & \multicolumn{3}{c}{MDB-Drums} & \multicolumn{3}{c}{RBMA13}  \\
\cmidrule(lr){4-6} \cmidrule(lr){7-9} \cmidrule(lr){10-12}
& & & micro & macro & macro* & micro & macro & macro* & micro & macro & macro*
\\
\midrule
All & Proto. Net & Include & 0.55  & 0.54  & 0.89  & 0.60 & 0.61 & 0.90  & 0.54  & 0.53 & 0.83 \\
&  & Exclude & 0.49  &  0.45 & 0.87  & 0.53 & 0.48 & 0.88  & 0.52  & 0.48 & 0.81 \\
\bottomrule
\end{tabular}
\caption{F-measure evaluated on real polyphonic music datasets under all classes that exist in each dataset.}
\label{tab:open-vocab}
\end{table*}


We use a backbone convolutional neural network (CNN) to embed the input as shown in \fref{fig:emb}. It consists of four convolution blocks, each of which has a convolutional layer with a $3 \times 3$ kernel, a batch normalization layer, a ReLU activation layer, and a $2 \times 2$ max-pooling layer. To allow our model to handle varying-duration input, we apply max-pooling along the time dimension to the output of the convolution blocks (rather than a fully connected layer). Finally, we flatten the feature map to get an embedding with a dimensionality of 1024. We train our model using the \texttt{Adam} optimizer \cite{kingma2014} in \texttt{PyTorch} \cite{paszke2019} with a learning rate of 0.001 for 100,000 episodes with early stopping. We choose the best model based on the few-shot classification loss on the validation set.

\subsection{Evaluation}
To evaluate our proposed few-shot drum transcription paradigm, we apply the prototypical network trained on Slakh2100 to perform drum transcription on three real music datasets. For each dataset, we first evaluate transcription under the fixed vocabulary scenario by mapping percussion instruments to a predefined 18-class vocabulary used in a state-of-the-art ADT system \cite{vogl2018} for comparison. Then, we transcribe all classes that exist in the test set, including those classes that do not exist in our training data, mimicking the open vocabulary scenario.

Given a target class and an audio track, we first preprocess the track into a series of overlapping query frames with a 250 ms window size, matching the context window used during training, and 10 ms hop size. To simulate human selections at inference time, we randomly sample 5 target examples from the track as positive examples. Then, we take all frames within the track as negative examples and predict each query frame as described in \sref{few-shot-dt}. We run 10 iterations of this prediction process to account for randomness and concatenate all predictions to compute performance metrics. We estimate target onset locations from the model output using the peak picking method described in \cite{bock2012}. A frame $n$ is selected as an onset if the corresponding output probability $p(n)$ meets the following criteria:
\begin{enumerate}
    \item $p(n)=max(p(n-w_{1}):p(n+w_{2}))$, 
    \item $p(n) \geq mean(p(n-w_{3}):p(n+w_{4})) + \delta$, 
    \item $n-n_{last \: onset}>w_{5}$, 
\end{enumerate}
where $\delta$ is a threshold parameter, $w_{1}$ to $w_{4}$ are sample offset values defining the windows for the $max$ and $mean$ functions, and $w_{5}$ is the minimum allowed number of samples between onsets. 

We divide each dataset used for evaluation into three splits for 3-fold cross-validation. For each fold, we tune the peak-picking parameters on the validation split using a randomized search with 1000 iterations, and perform drum transcription on the test split. Finally, we report the model performance averaged over the three test splits. For ENST-Drums, each split contains a different drummer. For MDB-Drums and RBMA13, we use the same splits as \cite{vogl2018}.

\subsection{Metrics}
We compute performance metrics by first using the \texttt{onset\_evaluation} function in \texttt{madmom} \cite{bock2016} to find matching onset locations with a 20 ms tolerance window. We then compute F-measure as the primary performance metric, using both \textit{micro} and \textit{macro} aggregation. For \textit{micro} F-measure, we aggregate all true positives (TP), false positives (FP), and false negatives (FN) over all classes and tracks in the entire dataset. For \textit{macro} F-measure, we first compute a track-level F-measure for each track by averaging all class-level F-measures in the track, and we then average over all track-level F-measures in the dataset to compute the final metric.

When computing \textit{macro} F-measure, if an instrument does not exist in a track and the ADT model under evaluation does not predict any corresponding positive labels, previous work defined its class-level F-measure to be 1 \cite{vogl2018}. This convention is informative for a standard supervised approach since the model may produce false positives for non-existing classes. However, a few-shot model would never predict non-existing classes since there are no positive examples to begin with, which is one of the advantages of the few-shot drum transcription paradigm. Therefore, when evaluating our few-shot model, it makes more sense to exclude non-existing classes in a track from the evaluation to avoid artificially inflating the \textit{macro} F-measure. For completeness and comparison to previous work, however, we report both variants, either excluding non-existing classes in a track (\textit{macro} F-measure) or following the convention of setting the F-measure for such classes to 1 (\textit{macro*} F-measure).

Given that our model requires five user-labeled examples from the test data in each iteration of prediction, we can compute the aforementioned performance metrics either including or excluding the user-labeled examples. The former represents the joint human-computer performance of the proposed paradigm that a user would experience for a track, while the latter represents our model's performance on strictly unlabeled data (i.e., the rest of the track). We report both variants for each metric described in the previous paragraph (labeled as Include/Exclude in the Tables). 

\section{Results}
\subsection{Fixed Vocabulary}
We first compare our few-shot drum transcription approach to a state-of-the-art CRNN model, which was trained on synthetic data and fine-tuned on real music data, under a fixed 18-class vocabulary \cite{vogl2018}. Note that while NMF-based methods can be considered more closely related to our approach, they require iterative optimization at test time and the determination of the non-negative rank for the decomposition process can be difficult. Most previous NMF-based ADT systems were evaluated on solo drum tracks and a small subset of percussive sound classes \cite{Wu2015, Lindsay2012, dittmar14}. Therefore, in this work, we choose the CRNN model as the baseline system and plan to compare our approach to NMF-based methods as part of future work.

In \tref{tab:fix-vocab} we present model performance on three real music datasets. From these results, we find three distinct insights applicable to the fixed vocabulary ADT tasks. First, the results show that our approach, a prototypical network trained on synthetic data with only five examples provided at inference time, gives comparable and in some cases better performance compared to previous, fully supervised state-of-the-art results.  Second, we see that our model performance is relatively stable across different datasets. Third, our proposed approach outperforms the supervised model on RBMA13 by a large margin, which is considered a difficult dataset with diverse drum sounds \cite{vogl2018}. For instance, snare drum sounds on different tracks in RBMA13 are very diverse and can sound very different. Standard supervised approaches typically struggle to generalize well for classes with high intra-class variation. However, while a percussive sound class may exhibit large intra-class variation across different tracks (e.g. different tracks may have very different snare drum sounds), it's often the case that such sound classes display far less intra-class variation within the same track (e.g., the same snare drum is used throughout a track). Since our few-shot model detects a sound class based on target examples from the same track, it is considerably more robust when it comes to intra-class variation, as evidenced by the quantitative results. This highlights the strength of the few-shot drum transcription paradigm which instead of aiming at generalization, aims for quick adaptation with minimal human input.


\begin{figure}[t]
\centering\includegraphics[trim={0.29cm 0.38cm 0.3cm 0.25cm},clip, width=\columnwidth]{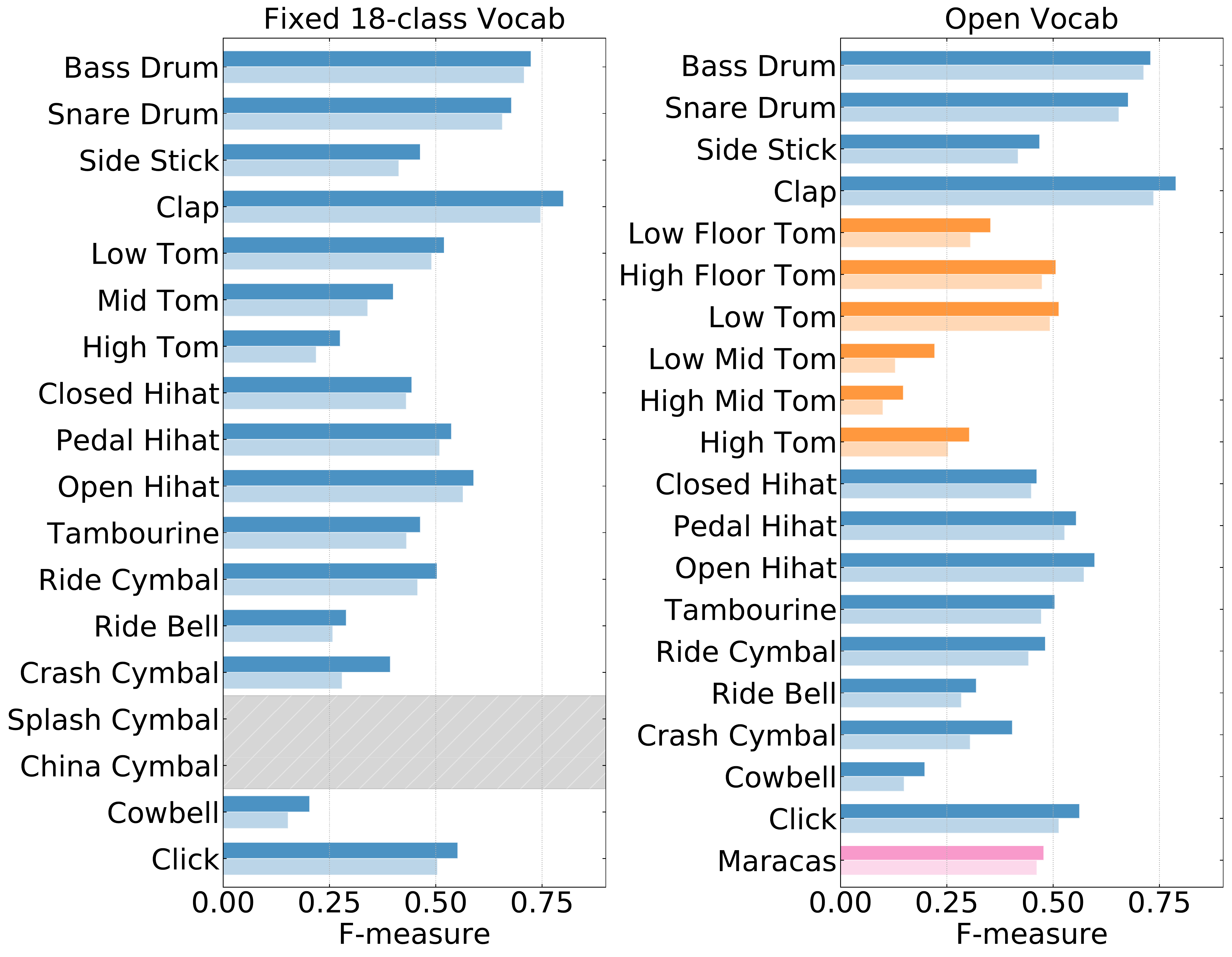}
\caption{F-measure for each percussion instrument in the RBMA13 dataset. (Left) Under 18-class vocabulary. (Right) Under all 21 classes. For each instrument, we show the metrics computed with target examples included (dark bar) and excluded (light bar).}
\label{fig:inst-break}
\end{figure}

\subsection{Open Vocabulary}
Next, we evaluate our few-shot model under an open-vocabulary scenario. Here, we evaluate the model against all the classes in each test set, including classes that were never seen by the model during training. Specifically, we evaluate on 20 classes for ENST-Drums, 19 classes for MDB-Drums, and 20 classes for RBMA13. Classes that do not appear more than five times in any track in the dataset are filtered out. Note that 6, 4, and 1 classes within the full vocabulary of ENST, MDB, and RBMA respectively do not exist in our training data. The results are presented in Table 2. Note that we do not compare our model to the fully supervised model in this scenario since, as noted earlier, such a model would fail (by design) to recognize classes that are outside of the training vocabulary. When we compare these results to those in \tref{tab:fix-vocab} (fixed vocabulary), we note that there is no drop in performance when moving from a fixed known vocabulary to an open vocabulary with previously unseen sound classes. This is a direct result of adapting few-shot learning to ADT and highlights the benefit of our proposed approach.

Next, we break down the performance of the few-shot model by instrument class under both the fixed and open vocabulary scenarios on RBMA13, presented in \fref{fig:inst-break}. Here, two out of 18 predefined classes, splash cymbal and china cymbal, do not exist in RBMA13 annotations and thus the results for these classes are absent from the figure. We see that in the open vocabulary scenario, we are able to transcribe fine-grained classes such as six different tom drums (orange bars in \fref{fig:inst-break}) with comparable performance to predicting a coarser, fixed vocabulary. We can also transcribe Maracas (pink bar in \fref{fig:inst-break}) which our model has not seen at training.

\subsection{Transcribing Novel Classes}
\label{sec:novel-class}
In the previous section, we saw that the model can detect a class that is out of the training vocabulary. To evaluate this more quantitatively, we re-train our few-shot model on the Slakh2100 dataset while completely excluding three classes from the training data: bass drum, tambourine, and clap, representing both common and rare classes. We then evaluate the model on predicting these three classes in the Slakh2100 test set and compare the results to those we obtain when the three classes are included in the training data.

We present the results in Table 3. We see that the performance of the model is very stable, with only a minor decrease in F-measure when predicting classes that are completely excluded from the training set. This confirms that with the few-shot training paradigm, our model can detect entirely unseen classes given just a few examples at inference time.

\begin{table}[t]
\centering
\small
\begin{tabular}{P{6em}P{4em}|P{2em}P{2em}P{2em}}
\toprule
Training data & Target examples & \multicolumn{3}{c}{$\mathcal{H}$ in Slakh2100 test set} \\
\cmidrule(lr){3-5}
 & & micro & macro & macro* \\
\midrule
Slakh2100 & Include & 0.66  & 0.66 & 0.83 \\
& Exclude & 0.64 & 0.64 & 0.83 \\
\cmidrule(lr){1-5}
Slakh2100 - $\mathcal{H}$  & Include & 0.62  & 0.62 & 0.81\\
 & Exclude  & 0.61  & 0.60 & 0.80 \\
\bottomrule
\end{tabular}
\caption{F-measure evaluated on three classes: $\mathcal{H} = \{\mbox{bass drum, tambourine, clap}\}$ in the Slakh2100 test set when training with the entire Slakh2100 training set or with three classes held out from the training data.}
\label{tab:slakh}
\end{table}



\subsection{Performance Breakdown by Polyphony}
Next, we focus on the target example selection process at inference time. Due to the polyphonic nature of music audio, when a target example is selected, it can include non-target instrument sounds, played at the same time. We want to investigate how the polyphony of these selected examples affects transcription performance. To do so, we repeat our evaluation process three more times, each time varying the support set such that all positive target examples have the same degree of polyphony: 1, 2, and 3 or more (3+). To define the polyphony of each example, we look at a 20 ms window around its onset and count the number of percussion instruments that co-occur within the window. To assess how the performance is affected by the polyphony of the \emph{query}, we break down the performance by the polyphony of the query frames.

\begin{figure}[t]
\centering
\includegraphics[trim={0cm 0cm 0cm 0cm},clip, width=\columnwidth]{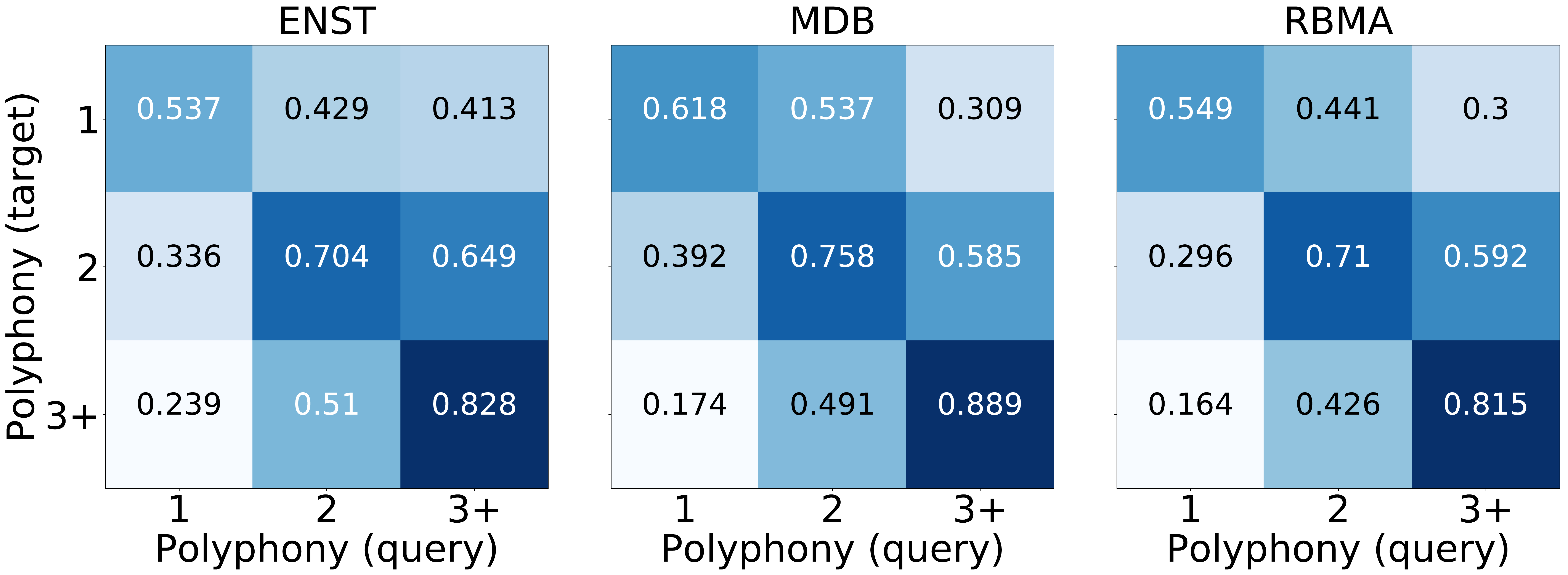}
\caption{Break down F-measure by polyphony in both target and query examples.}
\label{fig:poly}
\end{figure}

We present the results in \fref{fig:poly}, which show similar trends across the three evaluation datasets. First, we see high performance on the diagonal, where the polyphony between target and query examples match. Along the diagonal, performance also increases with increasing polyphony, for which a possible explanation is that the chance of having exactly matched instrument sources between target and query examples increases at high polyphony. That is, the number of different percussion instrument combinations decreases with increasing polyphony, due to the underlying pattern of drum playing. Another insight is that when there is a mismatch between target and query polyphony, having lower polyphony in target examples than in query examples gives better performance than the other way around (comparing the upper right triangle to the lower-left triangle in each figure). This matches the intuition that when the target examples have high polyphony, it is difficult for the few-shot model to latch onto the correct target instrument, resulting in poor performance even on query examples with low polyphony. On the other hand, it is easier for the model to find the target class in common within examples with lower polyphony.

Overall, the results show that the performance of our few-shot drum transcription approach can significantly depend on the target examples selected for the support set. In future work, we plan to build on top of these results to investigate the best strategy of composing a support set, and how we can inform user to make effective selections.

\section{Limitation and Future Work}
The main limitation of our proposed approach is that it requires the user to provide a few examples at inference time. 
If the vocabulary is known in advance and there is sufficient training data for each sound class, a fully supervised approach has the advantage of not requiring any user intervention. However, when the vocabulary is not known in advance, or when there isn't enough training data for every class, the few-shot paradigm is clearly advantageous. For future work, we plan to investigate the example selection process at inference time and the corresponding human element. This includes studying how we can compose sets of support examples to maximize performance, and how we can guide the user to those selections.

\section{Conclusion}
In this work, we address open vocabulary ADT by proposing a few-shot drum transcription paradigm, a combination of a few-shot model with minimal human input. We train a prototypical network on the Slakh2100 dataset as the few-shot model, 
and evaluate the proposed few-shot drum transcription system on multiple real-world ADT datasets with polyphonic accompaniment. We show that, given just a handful of target examples, we can match and, in some cases outperform, a state-of-the-art supervised ADT approach under a fixed vocabulary setting. At the same time, we show that our model can successfully generalize to finer-grained class labeling and extended vocabularies unseen during training. Lastly, we investigate the dependence of few-shot model performance on the polyphony of target examples. We show that matching polyphony in target and query examples gives better performance and when there is a mismatch, having lower polyphony in target examples than in query examples gives better results.   


\section{Acknowledgment}
The authors would like to thank Richard Vogl for sharing experimental details in the baseline system. 

\bibliography{ISMIRtemplate}


\end{document}